\documentclass[
 amsmath,amssymb,
 aps, prl, twocolumn, superscriptaddress
]{revtex4-2}

\usepackage{mathtools}
\usepackage{amsmath,amssymb}
\usepackage{graphics,graphicx,color, calc}
\usepackage{xcolor}
\usepackage{dcolumn}
\usepackage[english]{babel}
\usepackage{mathrsfs}
\usepackage[mathcal]{eucal}
\usepackage{bm}
\usepackage{lipsum}
\usepackage{xr}
\usepackage{stackengine}
\usepackage{xcolor}
\newlength\raisedepth
\def\[{\left[}
\def\]{\right]}
\def\({\left(}
\def\){\right)}

\usepackage{bookmark}

\newcommand{\jfi}
{\affiliation{James Franck Institute, The University of Chicago, Chicago, IL 60637}}
\newcommand{\rri}
{\affiliation{Soft Condensed Matter Group, Raman Research Institute, Bengaluru 560080, Karnataka, India}}

\begin{document}

\title{Kovacs-like memory effect in a sheared colloidal glass: role of non-affine flows}

\author{Maitri Mandal}
\rri

\author{Abhishek Ghadai}
\rri

\author{Rituparno Mandal}%
\thanks{These authors contributed equally to this work.}
\rri
\jfi

\author{Sayantan Majumdar}%
\thanks{These authors contributed equally to this work.}
\rri

 %& \noindent\large{Maitri Mandal~\textit{$^{a}$}, Abhishek Ghadai~\textit{$^{a}$}, Rituparno Mandal~\textit{$^{a\ b ~\ast}$}, Sayantan Majumdar~\textit{$^{a ~\ast}$}} \\%Author names go here instead of "Full name", etc.

%\includegraphics{head_foot/dates} & \noindent\normalsize{

\begin{abstract}

Memory effect reflects a system's ability to encode, retain and retrieve information about its past. Such effects are essentially an out-of-equilibrium phenomenon providing insight into the complex structural and dynamical behavior of the system. Kovacs effect is one such memory effect that is traditionally associated with thermal history. Although studies on the Kovacs-like memory effect have been extended to mechanical perturbations such as compression-decompression, whether such effects can also be observed under volume-conserving perturbations like shear, remains unclear. Combining experiments, simulations and linear response theory we demonstrate Kovacs-like memory effect in a sheared colloidal glass. Moreover, we explore the influence of non-linear perturbations and establish a correlation between the deviation from linear response prediction and microscopic non-affine flows generated due to such large deformations in affecting the memory effect. Our study not only extends Kovacs-like memory effect in the domain of volume-conserving mechanical perturbations, it also highlights the importance of the nature of underlying microscopic flows in controlling the memory effect in amorphous matter.

\end{abstract}

\maketitle

%\end{tabular}

 %\end{@twocolumnfalse} \vspace{0.6cm}

 % 
 %]
%%%END OF TITLE, AUTHORS AND ABSTRACT%%%

%%%FONT SETUP - please do not change any commands within this section
%\renewcommand*\rmdefault{bch}\normalfont\upshape
%\rmfamily
%\section*{}
%\vspace{-1cm}

%%%FOOTNOTES%%%

\footnotetext{\textit{$^{a}$~Soft Condensed Matter Group, Raman Research Institute, Bengaluru 560080, Karnataka, India}}
\footnotetext{\textit{$^{b}$~James Franck Institute, The University of Chicago, Chicago, IL 60637. }}
\footnotetext{\textit{$^{\ast}$~RM and SM contributed equally to this work.}}

%Please use \dag to cite the ESI in the main text of the article.
%If you article does not have ESI please remove the the \dag symbol from the title and the footnotetext below.
\footnotetext{\dag~Supplementary Information available: [details of any supplementary information available should be included here]. See DOI: 10.1039/cXsm00000x/}
%additional addresses can be cited as above using the lower-case letters, c, d, e... If all authors are from the same address, no letter is required

%\footnotetext{\ddag~Additional footnotes to the title and authors can be included \textit{e.g.}\ `Present address:' or `These authors contributed equally to this work' as above using the symbols: \ddag, \textsection, and \P. Please place the appropriate symbol next to the author's name and include a \texttt{\textbackslash footnotetext} entry in the the correct place in the list.}

%%%END OF FOOTNOTES%%%

%%%MAIN TEXT%%%%
\section{Introduction}

Memory refers to a system's ability to encode, maintain and recall information of its past. Exploring such memory effects not only helps gain deeper insights into the underlying physical principles of memory formation in a wide variety of systems but can also contribute to the design/ development of smart and functional materials~\cite{Keim2019}. Memory effects are fundamentally associated with the out-of-equilibrium behavior and can be of diverse types depending on the nature of the applied perturbations~\cite{Keim2019}. For instance, under fixed amplitude cyclic shear, amorphous systems can develop a memory of the amplitude of the applied deformation,~\cite{Chattopadhyay2022, Keim2022, Fiocco2014, Adhikari2018} or the perturbation direction~\cite{Gadala1980}. Return-point memory allows a system to recall specific states upon re-experiencing past conditions.~\cite{RPM} There are also examples of memory formation due to thermal history~\cite{kovacs1963glass,Ludovic,Bhattacharya,Mpemba_1969}. These memory effects are typically studied in systems with complex responses, such as a glass or a supercooled liquid in which logarithmic or stretched exponential relaxation is observed, in contrast to simple liquids or, Maxwellian visco-elstic systems that show simple exponential relaxation implying a well-defined relaxation time. Note, athermal systems can also show complex (power law) relaxation (via a mechanism called `athermal ageing' that has been pointed out very recently~\cite{S.Fielding}), leading to significant changes in the rheological response of athermal solids~\cite{Jack}.

A consequence of such complex relaxation process can lead to a striking phenomenon: the Kovacs memory effect~\cite{Mossa2004,Prados_2014,Lahini2017,Murphy2020,Mandal2021,Yu_2023}. In the original experiments describing the Kovacs effect,~\cite{kovacs1963glass} it was shown that when a system is quenched below its glass transition temperature and then slightly reheated, its volume evolution exhibits a non-monotonic behavior rather than monotonically progressing to the corresponding equilibrium value~\cite{Prados2010}. Later, it has been demonstrated that the Kovacs-like memory effect is not limited to thermal perturbations~\cite{Dillavou,Kursten,He,Morgan}.

For example, when a material is subjected to a step like mechanical load and subsequently the load is partially released after a waiting time, the corresponding normal force can exhibit non-monotonic evolution~\cite{Lahini2017,Murphy2020,Mandal2021}.
A more direct evidence of the memory can be found in the dependence of the peak position of this non-monotonic response on the waiting time~\cite{Lahini2017,Murphy2020,Mandal2021}. Therefore one can retrieve information about the past (the deformation history) of the system by analyzing this non-monotonic evolution, in particular by observing the peak position.
Although the Kovacs-like memory effect has been observed and analyzed quite extensively for a variety of systems under mechanical perturbations, such as compression and stretching~\cite{Lahini2017,Murphy2020,Dillavou,Kursten,He,Morgan}, these experiments do not have any control over the volume of the deformed sample. Hence, whether Kovacs like behavior can be observed under volume-preserving mechanical perturbations (such as shear) remains elusive. Additionally, beyond a successful application of linear response theory (LRT)~\cite{Mandal2021} to understand such effects for smaller perturbation, the role of non-linearity and microscopic flows in the Kovacs-like memory is still unclear.

These above-mentioned facets recognise the necessity to extend the exploration of Kovacs-like memory under shear perturbations. Understanding the role of shear could also provide deeper insights into the universality of the Kovacs-like memory effect across different mechanical deformation modes. We combine experiments on a colloidal glassy system, simulations of an amorphous harmonic solid and a linear response theory(LRT)~\cite{Mandal2021}, to investigate and understand the Kovacs-like memory effect under different step shear protocol. We demonstrate the Kovacs memory effect both in the experiment and in the simulation and also rationalize it using the linear response theory. We observe that the memory effect persists regardless of whether the perturbation is in the linear or in the non-linear regime. Boundary imaging reveals that non-affine deformation (microscopic flows) appears in case of larger (non-linear) perturbations and the mean non-affinity is closely correlated with the observed deviation from LRT prediction, across all perturbations in both the experiments and the simulations.

\section{Results and discussion}

\subsection{Two step shear and Kovacs memory}

\begin{figure}[t]
\centering
\hspace{-0.5cm}
  \includegraphics[height=10.85cm]{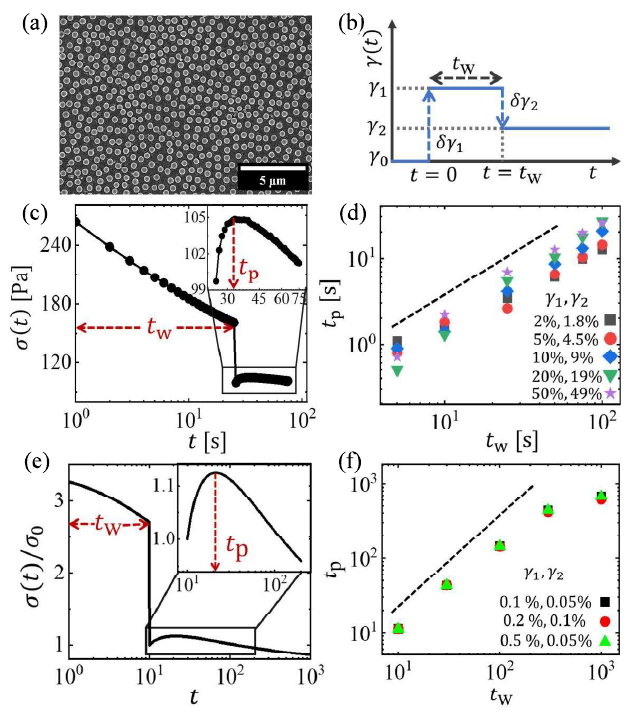}
  \caption{(a) SEM image of the PNIPAM particles (at lower volume fraction). (b) Schematic of the perturbation protocol used to study the Kovacs-like memory effect. (c) Stress response of the system in experiment, demonstrating Kovacs-like memory effect under shear (for $\gamma_1=10\%$, $\gamma_2=8\%$, $t_{\mathrm{w}}=25$ s). (d) Relation between the peak position of the stress response $t_{\mathrm{p}}$ and the waiting time $t_{\mathrm{w}}$, obtained from experiment. (e) Kovacs like non-monotonic stress response observed in simulation (with $\gamma_1=0.2\%$, $\gamma_2=0.1\%$, $t_{\mathrm{w}}=10$ s) and (f) a similar relationship between the peak position of the stress response, $t_{\mathrm{p}}$ and the waiting time $t_{\mathrm{w}}$, measured in simulations.}
  \label{fgr:protocol}
\end{figure}

We first synthesize soft, thermo-responsive PNIPAM particles (see Fig.~\ref{fgr:protocol}(a) for the SEM image of the synthesized particles) using one-pot method (see the Materials and methods section for details). For our rheological experiments, we use a dense aqueous suspension of these particles in a parameter regime where the system can practically be considered as a glassy solid and it exhibits extremely slow, logarithmic relaxation under deformation. In the later case the relaxation remains incomplete within the experimental timescale, with persistent residual stress, which is consistent with the glassy response of the system. All measurements are conducted at room temperature, using a stress-controlled rheometer (see the Materials and methods section for more details).

Fig.~\ref{fgr:protocol}(b) illustrates the experimental strain application protocol to study Kovacs memory effect. Initially, the system is at rest with zero strain at $\gamma_0=0$. We then apply a two-step shear strain perturbation. In the first step, a shear strain $\delta\gamma_1 = (\gamma_1-\gamma_0)$ is imposed and the system is held at that strain $\gamma_1$ for a specific waiting time $t_{\mathrm{w}}$. This is followed by a second shear strain $\delta\gamma_2 = (\gamma_1 -\gamma_2)$ in the opposite direction where the magnitude of $\gamma_2$ is smaller than $\gamma_1$. The system is held at the strain $\gamma_2$ for rest of the experiment. We performed our experiments for a wide range of $\gamma_1$ and $\gamma_2$. Additionally, for each pair of $\gamma_1$ and $\gamma_2$, the waiting time $t_{\mathrm{w}}$, is also varied within a range of 5 s to 100 s.

  The stress response of the system in our experiment, for the above mentioned protocol is depicted in Fig.~\ref{fgr:protocol}(c). Following the first-step perturbation, there is a sudden increase in stress due to the elastic response of the system. After this, the stress relaxes slowly in a logarithmic fashion. When the second-step strain is applied in the opposite direction, a similar elastic response causes a sudden drop in stress, followed by further relaxation. If we focus on the response of the system after the application of the second-step perturbation (as shown in the inset of Fig.~\ref{fgr:protocol}(c)), the non-monotonic stress relaxation becomes clearly visible. This non-monotonic behavior confirms the presence of the Kovacs-like memory effect under shear perturbation in our experiment. 

\begin{figure*}[t]
 \centering
 \includegraphics[height=12.9cm]{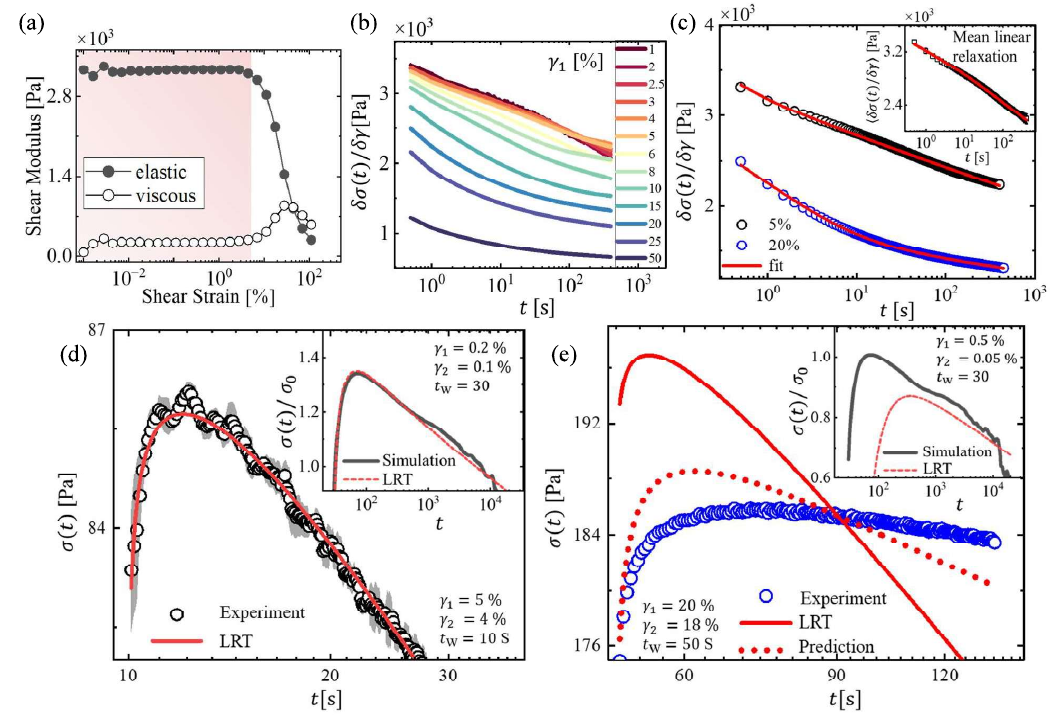}
 \caption{(a) Amplitude sweep measurement shows both the linear (until $\sim 5 \%$) and the non-linear ($> 5 \%$) regime. (b) Evolution of the instantaneous modulus ($\delta \sigma/\delta \gamma$) under step strain perturbation shows non-linear response for $\gamma_1 > 5 \%$. (c) Evolution of the instantaneous modulus for both small ($5 \%$) and large ($20 \%$) $\gamma_1$ fitted with double logarithmic $(a+ b \log(t)+ c \log(t+t_0))$ function. The inset shows mean of the instantaneous modulus for a set of linear perturbations ($\gamma_1$ between $1\%$ to $5\%$). (d) LRT prediction using the mean relaxation, shown in the inset of 2(c) (in red line) plotted along with the stress response obtained from the experiment. The black circles represent the experimental data, obtained using a 5-point moving average of the raw experimental data, and the shaded grey region represents the error bars, indicating the standard deviation of the averaged experimental data from the raw data. The inset shows a similar stress response in a two-step strain protocol in simulation along with the LRT prediction. (e) LRT prediction for the non-linear perturbation (solid red line) shows large deviations from experimental data (blue circles), whereas prediction using the relaxation function for $\gamma_1=20 \%$ provides a better fit. Inset shows a similar deviation (for large $\gamma_1$) between the stress response observed in simulation and the LRT prediction.}
 \label{fgr:Prediction}
\end{figure*}

The time at which the peak occurs after the second perturbation is referred to as the peak time, denoted by $t_{\mathrm{p}}$. According to the previous studies~\cite{Lahini2017,Murphy2020,Mandal2021}, the peak time ($t_{\mathrm{p}}$) linearly depends on the waiting time ($t_{\mathrm{w}}$), indicating that the system retains a memory of how long it is kept at the first step perturbation. In Fig.~\ref{fgr:protocol}(d), we plot $t_{\mathrm{p}}$ versus $t_{\mathrm{w}}$ for various $\gamma_1$, $\gamma_2$ values, including linear and non-linear perturbations (quantification of linear and non-linear perturbations is described in Sec.2.2). Fig.~\ref{fgr:protocol}(d) clearly shows that the $t_{\mathrm{p}}$ and $t_{\mathrm{w}}$ follow a linear relationship across a range of values of $\gamma_1$ and $\gamma_2$.
 
We also find Kovacs-like memory effect under shear in particle-based simulations of a sheared amorphous soft solid (see Fig.~\ref{fgr:protocol}(e)  and the Materials and Methods Section for simulation details). As before, in the case of simulation, we observe a non-monotonic response of the shear stress after the second step strain and observed the linear relationship between $t_{\mathrm{p}}$ and $ t_{\mathrm{w}}$ for different $\gamma_1$ and $\gamma_2$ as described in Fig.~\ref{fgr:protocol}(f). 

\begin{figure*}[t]
\centering
  \includegraphics[height=9.45cm]{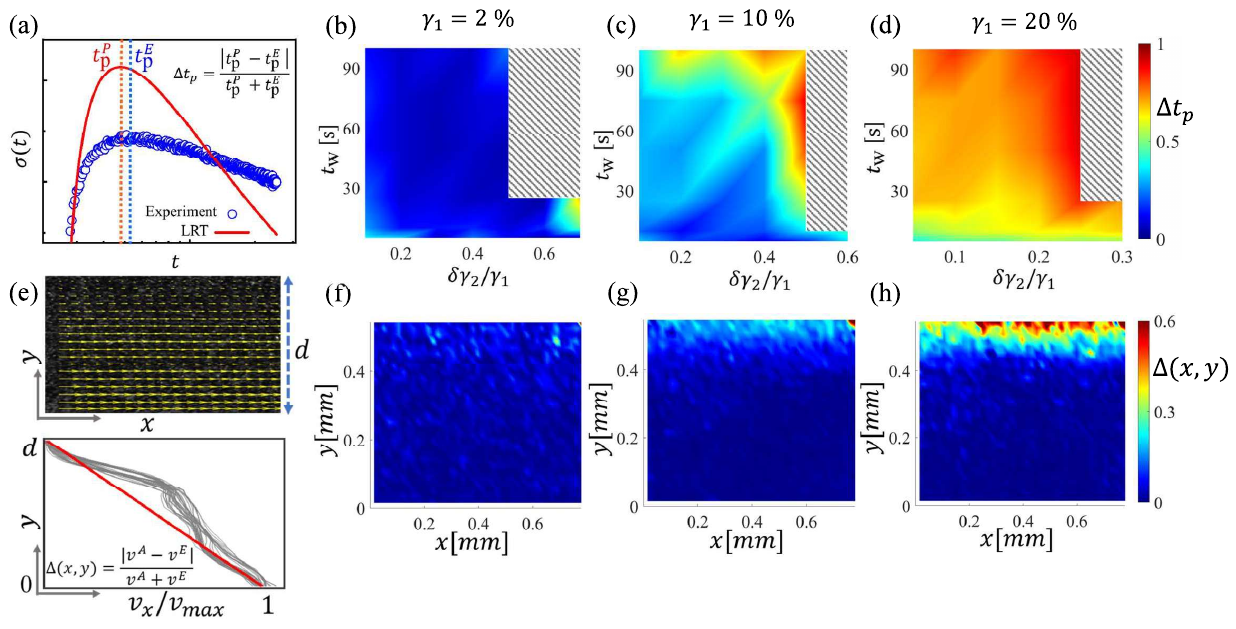}
   \caption{(a) Quantification of the deviation of the peak position as observed in experiments from the one predicted from LRT: $\Delta t_{\mathrm{p}}$. Predicted peak position: $t_{\mathrm{p}}^P$;  peak position obtained from experiments: $t_{\mathrm{p}}^E$. (b),(c) and (d) showing the dependence of $\Delta t_{\mathrm{p}}$ on $t_{\rm w}$ and $\delta \gamma_2/\gamma_1$ for increasing $\gamma_1$ ($2 \%$, $10 \%$ and $20 \%$). The shaded regions are state points where $t_{\mathrm{p}}^E$ could not be reliably extracted from experiments. (e) The top image shows the typical velocity profile obtained from the in-situ boundary imaging in our experiment. The gap between the top cone and the bottom plate is $d$; $x$, $y$ are the shearing and the velocity gradient direction respectively. In the bottom figure grey lines show the typical normalized velocity profile across the gap obtained at different $x$ coordinate. The red line represent the corresponding affine velocity profile.  $\Delta(x,y)$ is the measure of non-affinity; where $v^A(x,y)$ and $v^E(x,y)$ are the absolute values of affine and experimental velocities respectively (for definition see the SI S4). (f),(g) and (h) show the $\Delta(x,y)$ map for the corresponding $\gamma_1$ shown in subplot (b), (c) and (d), respectively.}
  \label{fgr:non_affinity_expt} 
\end{figure*}

\subsection{Linear response and the prediction}

Next, we use a linear response approach ~\cite{Mandal2021} to understand the stress response of our sheared amorphous system under Kovacs-like two step strain protocol. Linear response theory (LRT) provides a robust framework for studying memory effects in thermal~\cite{Prados2010} and athermal~\cite{Plata} systems. For a general time-dependent perturbation, the instantaneous response is expressed as 
\begin{equation}
\sigma(t) =  \int_0^t \chi(t - t') \dot{\gamma}(t')dt'
\label{lrt}
\end{equation}
where $\sigma(t)$ represents the response  of the system, $\gamma(t)$ is the applied time dependent perturbation, and $\chi(t)$ is the response function. It can be shown that $\chi(t)$ is system's response to a single-step perturbation and we measure $\chi(t)$ using this definition (both in experiment and simulation). Note $\sigma(t)$ in Eq.~\ref{lrt} describes the net change of stress in the system starting from an initial arbitrary value $\sigma(0)$.

 To proceed with the LRT formalism for Kovacs like the two-step strain protocol, we begin by asking the question `how large is the linear response regime?'. For this, we perform oscillatory shear experiments (amplitude sweep tests at a fixed frequency of $0.5$ Hz), as shown in Fig.~\ref{fgr:Prediction}(a). Additionally, we study the evolution of the instantaneous modulus (defined as $\delta\sigma/\delta\gamma$, where $\delta \sigma$ is the change in shear stress due to strain $\delta \gamma$) of the system as presented in Fig.~\ref{fgr:Prediction}(b). Both approaches indicate that the system response remains linear up to $\sim 5\%$ shear strain. In the case of simulations, a similar analysis yields a somewhat smaller linear response regime (up to $\gamma \sim 0.2\%$, as shown in {{the SI}} Fig.S1(a)).

\begin{figure*}[t]
\hspace{-0.30cm}
\centering
  \includegraphics[height=9.35cm]{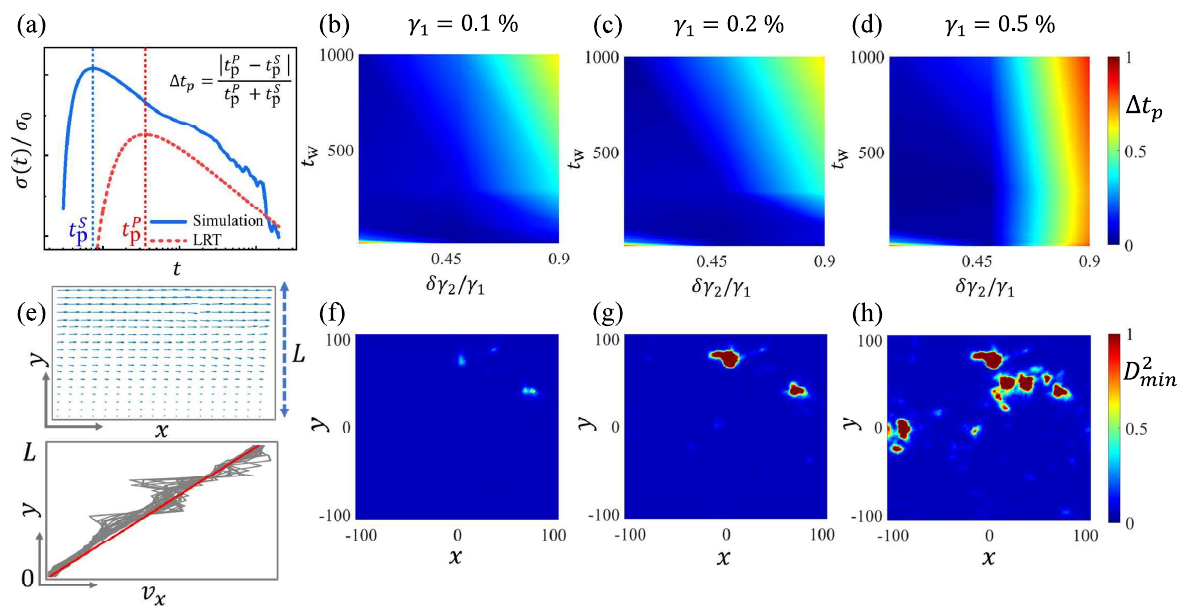}
   \caption{(a) Quantification of the deviation of the peak position as seen in simulations  from the one predicted from LRT: $\Delta t_{\mathrm{p}}$. Predicted peak position: $t_{\mathrm{p}}^P$;  peak position obtained from simulations: $t_{\mathrm{p}}^S$. (b),(c) and (d) showing the  dependence of $\Delta t_{\mathrm{p}}$ on $t_{\rm w}$ and $\delta \gamma_2/\gamma_1$ for increasing $\gamma_1$. (e) The top image shows the typical velocity vectors over the whole system. The shearing and the velocity gradient directions are respectively $x$ and $y$. In the bottom figure grey lines show the typical velocity profile across $y$, measured at different $x$ coordinate. The red line is the corresponding affine velocity profile. (f),(g) and (h) show the $D^2_{min}$ map for the corresponding $\gamma_1$ shown in subplot (b), (c) and (d) respectively.}
   \label{fgr:D2min_sim}
\end{figure*} 

Previous studies~\cite{AOI,Murphy2020} have explored stress relaxation due to single step deformation in disordered and glassy systems, suggesting that this relaxation behavior results from a wide distribution of relaxation times. This complex relaxation process is often well-described by a double logarithmic function~\cite{Lahini2017, Murphy2020, Mandal2021} which clearly indicates the absence of a single relaxation timescale. Fig.~\ref{fgr:Prediction}(c) shows the single-step stress relaxation response from experiment for perturbations both in the linear ($\gamma_1 = 5\%$) and the non-linear ($\gamma_1 = 20\%$) regime, each fitted with the same form of double logarithmic function:
\begin{equation}
\frac{\delta \sigma (t)}{\delta\gamma}=a+blog(t)+clog(t+t_0).
\label{fit}
\end{equation} 
This suggests that our system follows the same relaxation function both in the linear and in the non-linear regime, demonstrating that the relaxation behavior is unchanged across different types of perturbations (a similar set of data from the simulation is presented in {{the SI}} Fig.S1(b)). For the linear response calculation from now on we focus on the linear regime of the single-step deformation. In the inset, we plot the mean of all the relaxation functions acquired within the linear regime and we fit this mean relaxation function $\langle \delta \sigma (t)/ \delta\gamma \rangle$  with the the functional form given in Eq.~\ref{fit}. For the linear response prediction, we use $\chi(t)=\langle \delta \sigma (t)/ \delta\gamma\rangle$ where the fit yields the parameter values  $~a = 3341.508, ~ b = -304.460, ~c = -146.822$ and $~t_0 = 6.338$ corresponding to Eq.~\ref{fit}.

As our perturbation involves two steps, the deformation $\gamma(t)$ can be represented as a sum of two Heaviside theta functions (opposite in sign): 
\begin{equation}
\gamma(t) = \gamma_0+ \delta \gamma_1 \Theta(t) - \delta \gamma_2 \Theta(t - t_w)
\label{perturbation}
\end{equation} 
where $\gamma_0=0$, $\delta \gamma_1=\gamma_1-\gamma_0$ and $\delta \gamma_2=\gamma_1-\gamma_2$. Using LRT as described in Eq.~\ref{lrt} and combining with the definition of $\gamma(t)$ from Eq.~\ref{perturbation} we finally get the stress response,
\begin{equation}
\sigma(t) = \delta\gamma_1\chi(t)-\delta\gamma_2\chi(t-t_{\mathrm{w}}).
\label{prediction}
\end{equation}
To assess the predictability of LRT, we choose both $\gamma_1$ and $\gamma_2$ in the linear regime and compare the LRT prediction (from Eq.~\ref{prediction}) with corresponding two step strain experiment. For the data shown in Fig. ~\ref{fgr:Prediction}(d), we have chosen $\gamma_1=5\%$ and $\gamma_2=4\%$ (both $\delta\gamma_1$ and $\delta\gamma_2$ are in the linear regime) with an waiting time of 10 s. As shown in Fig.~\ref{fgr:Prediction}(d), we find a very good agreement between experimental data and LRT prediction. Here we note that the data in both the cases show the corresponding responses only after the second perturbation. A similar exercise in simulation (with $\gamma_1=0.2\%$ and $\gamma_2=0.1\%$;  again $\delta\gamma_1$ and $\delta\gamma_2$ both are in the linear regime) confirms the LRT prediction.

As discussed, we observe the Kovacs-like memory effect for linear, as well as, non-linear strain perturbations. The question that naturally arises now is how different the LRT prediction and the outcome of a Kovacs strain protocol will be beyond the linear regime. For this, we use $\gamma_1=20\%$ and $\gamma_2=18\%$ where only the first perturbation ($\delta \gamma_1=20 \%$) is far away from the linear regime, but the second one ($\delta \gamma_2=2 \%$) is well within it. A clear deviation from the LRT prediction can be observed in this case (see Fig.~\ref{fgr:Prediction}(e). On a similar line, the inset of Fig.~\ref{fgr:Prediction} (e), shows stress response measured from simulation along with the LRT prediction for larger perturbation. For larger perturbations, instead of using the linear response function $\chi(t)$, one can make a somewhat better prediction using $\frac{\delta \sigma}{\delta \gamma}$ obtained directly from the stress response corresponding to the non-linear strain perturbation$\delta \gamma_1=20 \%$ (Fig. ~\ref{fgr:Prediction} (e)). Nonetheless, considerable deviation from the LRT prediction is observed for such non-linear perturbations.
 In summary, for both simulations and experiments, the LRT predictions show good agreement with the experimental or simulation data, as long as the perturbations are in the linear regime.

\subsection{Quantification of the deviation from the LRT prediction and measurement of non-affinity}

Next, we quantify the deviation of LRT prediction for non-linear perturbations. For this, we investigate the difference between the peak positions obtained from the experiment and those predicted by LRT. We define a quantity, $\Delta t_{\mathrm{p}}$, which measures the absolute difference (see Fig.~\ref{fgr:non_affinity_expt}(a)) between the peak time $t_{\mathrm{p}}^E$ observed in the experiment and that predicted from LRT $t_{\mathrm{p}}^P$. 
\begin{equation}
\Delta t_{\mathrm{p}}=\frac{|t_{\mathrm{p}}^P-t_{\mathrm{p}}^E|}{t_{\mathrm{p}}^E+t_{\mathrm{p}}^P}
\end{equation}
where the denominator acts as a normalization constant. 
We then calculate $\Delta t_{\mathrm{p}}$ for a wide range of $\gamma_1$, $\gamma_2$ and $t_{\mathrm{w}}$. For comparison between different cases, $\Delta t_{\mathrm{p}}$ is plotted as a color map as a function of waiting time $t_{\mathrm{w}}$ and $\frac{\delta\gamma_2}{\gamma_1}$ which represents the magnitude of the second perturbation relative to that for the first step {\it{i.e.}} $\gamma_1$.

The experimental data suggest that for a fixed $\gamma_1$ the prediction deviates significantly with increasing $\frac{\delta \gamma_2}{\gamma_1}$ and also for higher $t_{\rm w}$, as highlighted by the increasing $\Delta t_{\mathrm{p}}$ values in Fig.~\ref{fgr:non_affinity_expt}(b) to ~\ref{fgr:non_affinity_expt}(d). For a fixed value of $\frac{\delta \gamma_2}{\gamma_1}$ and $t_{\rm w}$, the deviations become larger for increasing $\gamma_1$ as indicated in the figure. Similar data for $\gamma_1=5\%$ and $\gamma_1=50\%$ are shown in {{the SI}} Fig.S2. The shaded regions in the figure represent the parameter space where we do not obtain a clear peak in the experiments (see the SI Fig.S3 for details).

Until now, we obtain the deviation in the peak position for increasing $\gamma_1$, $\delta \gamma_2$ and $t_{\rm w}$. However, the connection between such behavior and the underlying flow properties of the system is not yet clear. In this section, we explore the bulk response of the system in light of its microscopic flow behavior. We perform in situ boundary imaging (imaging setup is reported in Ref.~\cite{chattopadhyay_2022}) to visualize such flows. Since the system is optically transparent, we added a small amount of tracer Polystyrene particles for visualization. We captured flow images during the first step perturbation $\delta{\gamma_1}$ (see the Materials and Methods section for further details) and analyzed the images using particle imaging velocimetry (PIV). The velocity vector map obtained using PIV across the entire region created by the gap $d$ ($\sim 0.5 mm$ between the top cone and the bottom plate of the rheometer) and a small section ($\sim 0.7 mm$) in the shearing direction $x$ is shown in the top panel of the Fig.~\ref{fgr:non_affinity_expt}(e). The figure clearly illustrates the flow generated due to the movement of the bottom plate and the resulting velocity gradient in the $y$-direction. The bottom panel of Fig.~\ref{fgr:non_affinity_expt}(e) shows the normalized velocity component along the  $x$-direction ($v_x/v_{\rm max}$) across the gap. The grey lines represent the experimental velocity profile (measured at different x-slices of width $\sim 0.02 mm$), while the red line shows the corresponding affine velocity profile. 

We measure the experimental velocity field ($v^E$),  estimate an affine flow field ($v^A$) for the same profile (for details see the SI S4), and then calculate non-affine field $\Delta(x,y)$ using these two.
The absolute difference between $v^E$ and $v^A$, normalized by their sum provides a measure of non-affinity  $\Delta(x, y) = \frac{\left| v^A - v^E \right|}{v^A + v^E}$. Fig.~\ref{fgr:non_affinity_expt}(f) to ~\ref{fgr:non_affinity_expt}(h) show the corresponding non-affinity maps $\Delta(x,y)$ for increasing shear strain (single step $\gamma_1=2 \%, 10 \%$ and $20 \%$). When the perturbation ($\gamma_1$) is small, the map shows that the velocity profile is almost entirely affine as represented by the dark blue color (see Fig.~\ref{fgr:non_affinity_expt}(f)). For $\gamma_1 =10\%$, the non-affinity gets slightly stronger. When $\gamma_1$ is increased to $20\%$ non-affinity appears in a banded pattern near the top cone, as indicated by red and yellow regions in the map. The non-affinity maps for the other strain values $\gamma_1=5\%$ and $\gamma_1=50\%$ are shown in SI Fig. S4 (a) and (b), respectively. 

In our simulation, we quantified the deviation from the LRT in the same way as described before as shown in ~\ref{fgr:D2min_sim}(a). In Fig.~\ref{fgr:D2min_sim}(b) to ~\ref{fgr:D2min_sim}(d), we show the normalized  deviation of the peak time $\Delta t^{}_p$ for different $t_{\rm w}, \delta \gamma_2/ \gamma_1$  and $\gamma_1$. At longer waiting times $t_{\rm w}$, larger $\gamma_1$ and $\delta \gamma_2/\gamma_1$ values, deviations becomes progressively visible. The top panel of Fig.~\ref{fgr:D2min_sim}(e) shows the velocity map of the particles, while the bottom panel displays a set of typical velocity profiles. To calculate the non-affinity in the simulation, we calculate $D^2_{min}$ following Ref.~\cite{Falk}. The $D^2_{min}(x,y)$ spatial maps for the corresponding single step deformation $\gamma_1$ are shown from Fig.~\ref{fgr:D2min_sim}(f) to ~\ref{fgr:D2min_sim}(h). In our simulation, we find that non-affinity arises locally and increases systematically with increasing $\gamma_1$. 

\begin{figure}[t]
\centering
  \includegraphics[height=11.5cm]{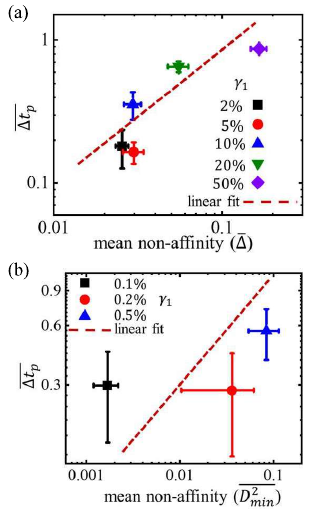}
   \caption{(a) Correlation between the mean of the deviation of the peak time  ($\overline{\Delta t_{\mathrm{p}}}$) and the mean non-affinity ($\overline{\Delta}$), obtained from multiple experiments performed at $\gamma_1$ in the range between ($2 \% - 50 \%$). (b) Correlation between the mean of the deviation of the peak time $\overline{\Delta t_{\mathrm{p}}}$ and the mean non-affinity ($\overline{D^2_{min}}$), extracted from different simulations performed at $\gamma_1$ in the range between ($0.1 \% - 0.5 \%$). In both cases (a and b), the error bars represent the standard deviation of the data obtained for different waiting times ($t_{\mathrm{w}}$) and strain ratios ($\delta\gamma_2/\gamma_1$) for each initial perturbation magnitude $\gamma_1$.
}
   \label{fgr:correlation}
\end{figure} 

\subsection{ Correlation between the deviation from LRT and the non-affinity}

In both experiments and simulations we observe that with increasing $\gamma_1$ the deviation of the predicted peak position and the overall non-affinity in the system increases. To investigate this further, for each $\gamma_1$, we define the mean deviation of the peak time as $\bar{\Delta t_p}$, the average of $\Delta t_p$ over the parameter space of $\{ t_{\rm w}, \delta \gamma_2/\gamma_1 \}$
and the mean non-affinity as,
\begin{equation}
\bar{\Delta}= \frac{1}{A}\int_x \int_y \Delta (x,y) dx dy= \frac{1}{A} \int_x \int_y  \frac{\left| v^A - v^E \right|}{v^A + v^E} dx dy
\label{non-affinity}
\end{equation}
where $A$ is the area of the system over which we analyze the velocity field.
To explore the relationship between the mean deviation of the peak time $\bar{\Delta t_p}$ and the mean non-affinity $\bar{\Delta}$, we show scatter plots in Fig.~\ref{fgr:correlation}(a) (for experiment) and Fig.~\ref{fgr:correlation}(b) (for simulation). The data indicates a significant correlation between the deviation of LRT and the average non-affinity present in the system.

\section{Conclusion}

In conclusion, our study demonstrate a Kovacs-like memory effect under shear perturbations in a colloidal glassy system. Simulation of a model soft amorphous system and linear response theory, also capture such interesting phenomena. Both experimental and simulation results demonstrate that the linear relationship between the peak time ($t_{\mathrm{p}}$) and the waiting time ($t_{\mathrm{w}}$) persists across a broad range of shear strains, even including those beyond the linear regime. It is not clear why such linearity between $t_{\mathrm{p}}$ and $t_{\mathrm{w}}$ also holds even deep inside the non-linear deformation regime and remains an interesting future direction to explore. We observe a gradually stronger deviation from the LRT prediction as we go deeper into the non-linear flow regime. Using in-situ boundary imaging, we quantify the non-affinity created for different strengths of ($\delta{\gamma_1}$).
We also quantify the non-affinity from our simulations. Interestingly, the nature of non-affinity is very different in experiments and simulations: in experiments, we obtain a band-like nonaffinity, whereas, in simulations, non-affinity appears as localized patches. 
Nonetheless, $t_{\mathrm{p}} - t_{\mathrm{w}}$ linearity holds in both cases irrespective of the nature of the non-affinity. 
For different strengths of the first-step perturbation, we calculate the mean deviation of the peak time and the average non-affinity. 
Despite the difference in the detailed nature of nonaffine deformations, a clear correlation between the mean deviation of peak time and the average non-affinity is obtained for both the experimental and the simulation, across the wide perturbation range explored in our study.
Obtaining an insight into the microscopic mechanism governing the nature of non-affinity (band-like or localized) constitutes another important future avenue to explore.

\section{Materials and methods}
\subsection{Experiment}

For experiments, PNIPAM particles are synthesized via one-pot free radical polymerization~\cite{brijitta} method. The microgel suspension underwent purification through repeated cycles of centrifugation, decantation, and redispersion. Subsequently, the PNIPAM particles were characterized using dynamic light scattering (DLS) and scanning electron microscope(SEM) imaging. At 25$^{\circ}$C, the particles have a mean diameter $d_h = 0.97 \pm 0.065 \, \mu\text{m}$, with the Volume Phase Transition (VPT) temperature approx at 33$^{\circ}$C~\cite{shibil}. A dense %(\blue{how dense? can we put some number density?}) 
aqueous suspension of these PNIPAM particles is used for rheological measurements. Rheological measurements were conducted using an MCR-702 stress-controlled rheometer (Anton Paar, Graz, Austria) equipped with a sandblasted cone-plate geometry (diameter: 25 mm, cone angle: 2$^{\circ}$, truncation height: 106 $\mu$m, gap: 0.5 mm) and a peltier temperature-controlled bottom plate. The rough, sandblasted surfaces of the measuring geometry minimize the wall-slippage during the rheological measurements. All experiments were performed at room temperature (25$^{\circ}$C). As the suspensions of PNIPAM particles are optically quite transparent, we introduce 1 wt\% polystyrene microspheres with a diameter of approximately 3.34  $\mu$m into the suspension to enhance the scattering. For in-situ boundary imaging, a high-speed imaging system is integrated with the rheometer, which operates in a separate motor transducer (SMT) mode. During the rheological tests, the sample boundary is illuminated using an LED light source (Dolan-Jenner Industries), and the scattered light is captured in the flow-gradient plane by a high-speed monochrome CMOS camera (Phantom Miro C210) equipped with a 10X long working distance objective (Mitutoyo). The flow profile is then mapped out from the boundary images using the Particle Image Velocimetry (PIV) technique. We used MATLAB software for PIV analysis, employing PIV codes developed by Nobuhito et al.(\url{https://www.mathworks.com/matlabcentral/fileexchange/2411-mpiv}).

\subsection{Simulation}

For our numerical simulation we employed a model athermal system ~\cite{Durian_PRL,Durian_PRE,S.Fielding, Mandal2021} comprising of soft bidisperse (equal numbers of type 1 and type 2) particles with radii respectively $R_1$ and $R_2$ where $R_1 = R$ and $R_2 = 1.4R$. The particles interact via repulsive harmonic pairwise interaction:
\begin{equation}
V_{ij}(r) = \frac{\kappa}{2} R_0^3 \left[ 1 - \frac{r}{\Sigma_{ij}} \right]^2 \Theta(\Sigma_{ij} - r)
\end{equation}
where $\Sigma_{ij} = R(i) + R(j)$ is the sum of the particle radii (where $R(i)$ refers to the $i$ th particle's radius that can be either $R_1$ or $R_2$ depending on the particle type), $r$ is the interparticle distance and $\Theta(y)$ is the Heaviside step function of variable $y$. Here $\kappa$ and $R_0$ controls the overall energy scale and we put them to unity. We use $N = 10^4$ particles ($N_1=N_2=5000$) in a undeformed box of linear size $L=216$ in the athermal limit ($T=0$).

Shear perturbations are implemented by imposing time-dependent strains $\gamma(t)$ (step like in nature for this study). Time evolution of the system is carried out using Lees-Edwards~\cite{Lees} periodic boundary conditions in a sheared geometry. The dynamics of the particles are modeled using an overdamped equation of motion:
\begin{equation}
\nonumber
\frac{d\mathbf{r}_i}{dt} = -\frac{1}{\eta} \nabla_i \sum_{j \neq i} V(|\mathbf{r}_i - \mathbf{r}_j|)
\end{equation}

where, $|\mathbf{r}_i - \mathbf{r}_j|=r$ is the inter-particle distance, $\mathbf{r}_i$ is the position vector of the particle $i$, and $\eta$ is the drag coefficient. By setting $\eta=1$, the characteristic timescale becomes
$\tau_0 = \frac{\eta}{\kappa R_0} = 1$. The equations of motion are evolved using an Euler integration scheme with a fixed time step $\delta t=0.01$. The packing fraction $\phi$ is defined as:

\begin{equation}
\nonumber
\phi = \frac{L^2}{\pi} \left( N_1 R_1^2 + N_2 R_2^2 \right)
\end{equation}

where $N_1=N_2=N/2$. Initial configurations are prepared by randomly distributing particles in the simulation box at $\phi=1$. The system is relaxed for $t=10^6$ to reach an energy-minimized state. These relaxed configurations are then subjected to different shear perturbations (one-step, two-step etc.). All the results presented here are averaged over 32 independent simulation runs that start from independent initial conditions. 
 
\section{Author contributions}

SM, MM, and AG conceived the project; MM and AG performed the experiments; RM did the simulations; MM, RM, and SM analysed data and wrote the paper.

%We strongly encourage authors to include author contributions and recommend using \href{https://casrai.org/credit/}{CRediT} for standardised contribution descriptions. Please refer to our general \href{https://www.rsc.org/journals-books-databases/journal-authors-reviewers/author-responsibilities/}{author guidelines} for more information about authorship.

\section{Conflicts of interest}

%In accordance with our policy on \href{https://www.rsc.org/journals-books-databases/journal-authors-reviewers/author-responsibilities/#code-of-conduct}{Conflicts of interest} please ensure that a conflicts of interest statement is included in your manuscript here.  Please note that this statement is required for all submitted manuscripts.  If no conflicts exist, please state that ``
There are no conflicts to declare.

\section{Data availability}

The data that support the findings of this study are available
from the corresponding author upon reasonable request. 
%A data availability statement (DAS) is required to be submitted alongside all articles. Please read our \href{https://www.rsc.org/journals-books-databases/author-and-reviewer-hub/authors-information/prepare-and-format/data-sharing/#dataavailabilitystatements}{full guidance on data availability statements} for more details and examples of suitable statements you can use.

\section{Acknowledgements}

SM acknowledges SERB (DST, Govt. of India) for support through a Ramanujan Fellowship and Raman Research Institute for intramural support.
We acknowledge Ranjini Bandyopadhyay for granting access to her lab facilities and  K.M. Yatheendran for help with the SEM imaging.

%%%END OF MAIN TEXT%%%


%\balance


%%%REFERENCES%%%
\bibliography{kovacs} %You need to replace "rsc" on this line with the name of your .bib file
\bibliographystyle{kovacs} %the RSC's .bst file

\clearpage

\section*{Supplementary Information}
\maketitle
\setcounter{figure}{0}
\renewcommand{\thefigure}{S\arabic{figure}}
\setcounter{section}{0}
\renewcommand{\thesection}{S\arabic{section}}

\section{S1 Estimate of linear response regime in simulation:}

\begin{figure*}[h]
   \begin{center}
     
        \includegraphics[height=6.5cm]{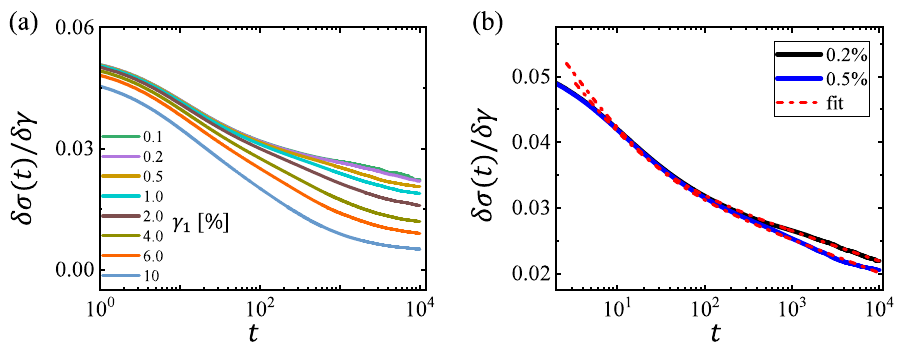}
        \caption{(a) Evolution of the instantaneous modulus $\frac{\delta \sigma (t)}{\delta\gamma}$ for a range of single step perturbation $\gamma_1=0.1 \% - 10 \%$ indicates the linear regime in simulation. (b) Instantaneous modulus for single-step perturbation is fitted with double logarithmic $\frac{\delta \sigma (t)}{\delta\gamma}=a+b \log(t)+c \log(t+t_0)$ form in both the linear ($\gamma_1=0.2$) and non-linear ($\gamma_1=0.5$) regime (fitting parameters are described in the text). }
        \label{relaxation}
    \end{center}
\end{figure*}

As mentioned in the main text, to measure the response under single-step perturbation in the simulation, we monitor the evolution of the stress $\sigma(t)$ just after the application of the perturbation. From this we calculate instantaneous modulus, $\delta\sigma(t)/\delta\gamma=\frac{\sigma(t)-\sigma(0)}{\gamma_1-\gamma_0}$ and we have plotted the modulus in Fig.~\ref{relaxation}(a). The evolution of the modulus seem to overlap with each other for a step strain ($\gamma_1$) of 0.2\% or below; beyond this point, the curves deviate, indicating the onset of nonlinearity.

In Fig.~\ref{relaxation}(b), we have shown instantaneous modulus for two different step strain $\gamma_1$: one in the linear regime ($\gamma_1=0.2\%$) and the other beyond the linear regime ($\gamma_1=0.5\%$). Similar to what we have observed in experiments, both the curves can be  fitted nicely with the same double logarithmic functional form: $\frac{\delta \sigma (t)}{\delta\gamma}=a+b \log(t)+c \log(t+t_0)$ where $a=0.03989$, $b=-0.01886$, $c=0.01436$, $t_0=17.2831$ for $\gamma_1=0.2 \%$ and $a=0.03987$, $b=-0.01977$, $c=0.01483$, $t_0=21$ for $\gamma_1=0.5 \%$.

\section{S2 Measurement of the peak time deviation:}

\begin{figure*}[h]
  \centering
   \includegraphics[height=6.0cm]{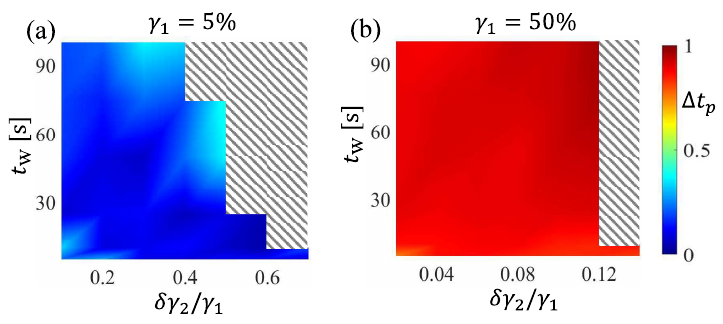}
        \caption{(a) and (b) showing the peak time deviations $\Delta t_{\mathrm{p}}$  for  $\gamma_1 = 5\%$ and $\gamma_1 = 50\%$ respectively. The shaded regions where $t_{\mathrm{p}}^E$ are absent.}
        \label{deviation}
\end{figure*}

To measure the deviation of the peak time $\Delta{t_{\mathrm{p}}}$ from LRT prediction, in both experiments and simulations, we determine the experimental peak time $t_p^E$ and the simulated peak time $t_{\mathrm{p}}^S$ from each of the two-step measurements. We have also predicted the peak time $t_{\mathrm{p}}^P$ using the linear response theorem in the context of two step strain protocol. Finally we define $\Delta  t_{\mathrm{p}}$ as the normalized difference between these two: $\Delta{t_{\mathrm{p}}} = \frac{|t_{\mathrm{p}}^P-t_{\mathrm{p}}^E|}{t_{\mathrm{p}}^E+t_{\mathrm{p}}^P}$ in experiments, and  $\Delta{t_{\mathrm{p}}} = \frac{|t_{\mathrm{p}}^P-t_{\mathrm{p}}^S|}{t_{\mathrm{p}}^S+t_{\mathrm{p}}^P}$ in the case of simulations.

We have shown this quantity $\Delta{t_{\mathrm{p}}}$ in Fig.3-4 of the main text. Here in Fig.~\ref{deviation}(a) and (b) we show the deviation of the peak time ($\Delta{t_{\mathrm{p}}}$), measured in experiments, for $\gamma_1=5 \%$ and $\gamma_1=50 \%$ respectively. Please note that the $\delta \gamma_2/\gamma_1$ range is not the same for $\gamma_1=5 \%$ and $\gamma_1=50 \%$ and we discuss this in detail in the next section.

%\section{S3 Absence of peak in the stress response for for larger $t_{\mathrm{w}}$ and $\delta\gamma_2$ :}

\section{S3 Absence of peak in the stress response for larger \texorpdfstring{$t_{\mathrm{w}}$ and $\delta\gamma_2$}{tw and delta-gamma2} :}

As both $t_{\rm w}$ and $\delta\gamma_2/\gamma_1$ increases, the parameter space over which we can reliably obtain the peak, gets smaller. The stress response for large $t_{\rm w}$ and large $\delta \gamma_2/\gamma_1$ gets flatter and it becomes progressively difficult to locate the peak and extract the peak position. This is the reason why the $\delta \gamma_2/\gamma_1$ range is not the same for $\gamma_1=5 \%$ and $\gamma_1=50 \%$, as shown in  Fig.~\ref{deviation}(a) and (b).

\begin{figure*}[h]
    \centering
   \includegraphics[height=11cm]{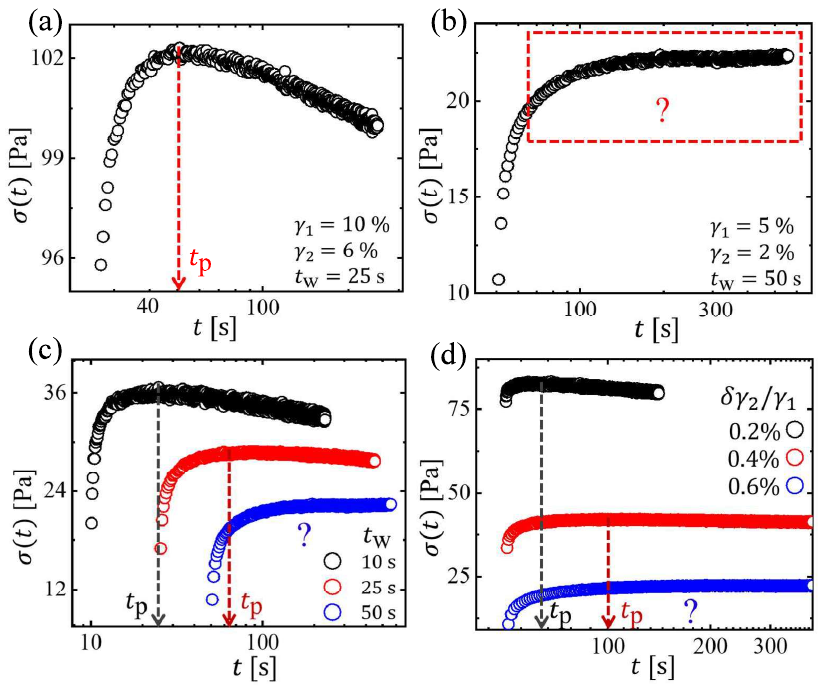}
        \caption{a typical example from experiment where (a) $t_{\mathrm{p}}$ is present (b) $t_{\mathrm{p}}$ is absent. (c) with increasing $t_{\mathrm{w}}$ getting $t_{\mathrm{p}}$ becomes harder (data shown for the fixed $\gamma_1 = 5\%$ and $\gamma_2 = 2\%$). (d) For the fixed $\gamma_1 = 5\%$ and $t_{\mathrm{w}}$ getting $t_{\mathrm{p}}$ of increasing $\delta\gamma_2/\gamma_1$ becomes difficult. }
        \label{tp_absent}
\end{figure*}

In Fig.~\ref{tp_absent}(a) (for $\gamma_1= 10\%,\ \gamma_2 = 6\%$  and $t_{\mathrm{w}} = 25s$), we show a typical non-monotonic stress response from the experiment, from which the peak position $t_p$ can be identified very easily whereas, for  Fig.~\ref{tp_absent}(b) (for $\gamma_1= 5\%,\ \gamma_2 = 2\%$  and $t_{\mathrm{w}} = 50s$), there is no clear peak in the stress response. In Fig.~\ref{tp_absent}(c) it has been demonstrated ( $\gamma_1 = 5\%$ and $\gamma_2 = 2\%$), that with increasing waiting time $t_{\rm w}$ it becomes harder to identify a clear peak and the corresponding peak time. Similarly in Fig.~\ref{tp_absent}(d) we show, for $\gamma_1$ (5\%) and ${t_{\mathrm{w}}}$ ($50 s$), the typical stress responses when $\delta\gamma_2/\gamma_1$ increases.

\section{S4 Measurement of the non-affinity in the experiment:}
\begin{figure*}[h]
  \centering
   \includegraphics[height=6cm]{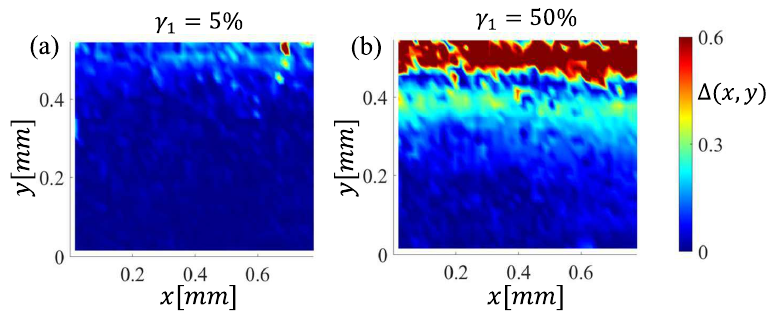}
        \caption{The non-affinity map $\Delta(x,y)$ for  (a) $\gamma_1 = 5\%$ and (b) $\gamma_1 = 50\%$ respectively, where $x$ is the shear direction and $y$ is the shear gradient direction. }
        \label{non-affinity_SI}
\end{figure*}

To calculate the non-affinity, we first measure the experimental velocity field $\mathbf{v}^E(x,y)$ and look at the absolute value of the normalised $x$-component velocity (where $x$ is the shear direction):
$v^E(x, y) = \left| \frac{v_x^E(x, y)}{v_{{max}}} \right|$
where, $v_{max}$ represents the maximum value of the x-component of $\mathbf{v}^E(x,y)$.
The experimental velocity field is averaged over different grid points along $x$-axis at two extremes of the profile, {\it{i.e.}} at $y = d$ and $y = 0$, yielding $\langle v^E(d) \rangle_x$ and $\langle v^E(0) \rangle_x$, respectively. Using these mean values, an affine flow field can be computed as:
$v^A(x,y) = \left[\frac{\langle v^E(d) \rangle_x - \langle v^E(0) \rangle_x}{d}\right]y + \langle v^E(0) \rangle_x$.
Note that by construction $v^A(x,y)$ depends only on $y$. Using the experimentally obtained velocity field $v^E(x,y)$ and the corresponding affine velocity field $v^A(x,y)$, we define a measure of non-affinity as:
$\Delta(x, y) = \frac{\left| v^A(x, y) - v^E(x, y) \right|}{v^A(x, y) + v^E(x, y)}$
where the denominator acts as a normalization constant. Non-affinity maps $\Delta(x, y)$ for $\gamma_1=5\%$ and $\gamma_1=50\%$ are shown in Fig.~\ref{non-affinity_SI} (a) and (b) respectively.

\end{document}